\documentclass[traditabstract]{aa} 
\usepackage{amsmath}
\usepackage{graphicx}
\usepackage{txfonts}
\usepackage{comment}
\usepackage{natbib}
\bibpunct{(}{)}{;}{a}{}{,}
\usepackage{float}
\usepackage{calc}
\usepackage{textcomp}
\usepackage{placeins}
\usepackage{color}
\usepackage{rotating} 
\usepackage{lscape}
\usepackage{url}
\usepackage[colorlinks=true,linkcolor=blue,citecolor=blue]{hyperref}
\usepackage{subfig}
\usepackage[all]{draftcopy}
\usepackage{longtable}
\usepackage{array}
\usepackage{threeparttable}
\usepackage{lscape}
\usepackage{enumerate}
\DeclareTextSymbol{\degre}{OT1}{23}

\newcounter{savedfootnote}

\renewcommand{\epsilon}{\varepsilon} 

\begin{document}
\title{Tracing early galaxies back in time: implications for the dearth of galaxies beyond $z\sim15$}

\author{
L.~Ciesla\inst{1}\fnmsep\thanks{\email{laure.ciesla@lam.fr}},
C.~Carvajal-Bohorquez\inst{1},
D.~Elbaz\inst{2},
G.~Aufort\inst{3},
O.~Ilbert\inst{1},
S.~Finkelstein\inst{4},
A.~M.~Koekemoer\inst{5},
P.~Oesch\inst{6,7},
M.~Boquien\inst{8},
C.~Gruppioni\inst{9},
N.~Laporte\inst{1}
}

\institute{	
Aix Marseille Univ, CNRS, CNES, LAM, Marseille, France
\and
Université Paris-Saclay, Université Paris Cité, CEA, CNRS, AIM, 91191, Gif-sur-Yvette, France
\and
Department of Astronomy, University of California, Berkeley, CA 94720, USA
\and
Department of Astronomy, The University of Texas at Austin, Austin, TX, USA
\and
Space Telescope Science Institute, 3700 San Martin Drive, Baltimore, MD 21218, USA
\and
Department of Astronomy, University of Geneva, Chemin Pegasi 51, CH-1290 Versoix, Switzerland
\and
Cosmic Dawn Center (DAWN), Niels Bohr Institute, University of Copenhagen, Jagtvej 128, DK-2200 København N, Denmark
\and
Université Côte d'Azur, Observatoire de la Côte d'Azur, CNRS, Laboratoire Lagrange, Nice, France
\and
Istituto Nazionale di Astrofisica - Osservatorio di Astrofisica e Scienza dello Spazio, Bologna, Italy
}

   \date{Received; accepted}

  \abstract
{
Although the \textit{James Webb} Space Telescope has pushed the observational frontier to $z\sim14$, the onset of star formation in the earliest galaxies remains beyond direct reach, making the star formation histories (SFH) of the galaxies we do observe a key tool for probing earlier epochs.
We aim to reconstruct the mass assembly histories of $z>8$ galaxies and infer their properties at earlier epochs.
For this, we develop and validate a new SFH model capable of reproducing the bursty star formation activity expected in high-redshift galaxies, based on an Ornstein-Uhlenbeck process.
Tested on the publicly available \textsc{sphinx$^{20}$} mock catalogs, the model provides accurate estimates of stellar masses and SFRs and reconstructs cumulative SFHs at population-level.
The earliest part of the SFH is not well recovered for galaxies at $z\sim8$ while the formation redshift is better constrained for galaxies at $z\gtrsim9$.
Applied to the UNICORN/GOODS-South sample, fitting 400 galaxies at $9\leq z\leq15$ to reconstruct galaxies' SFH, we find that sources observed at $z<10$ typically reached 20\% of their stellar mass no earlier than $z\sim12$, disfavoring a clear evolutionary link to spectroscopically confirmed galaxies at $z\gtrsim13$. 
Among 218 galaxies at $10<z<15$, only 28 had formed 20\% of their stellar mass by $z=15$, with a median formation redshift $z_{\rm form,20\%}=16.9^{+1.1}_{-1.1}$ ($z_{\rm form,50\%}=14.5^{+0.6}_{-0.7}$), and 24\% of our sample has already reached a stellar mass of 10$^7$\,M$_{\odot}$ by $z=15$.
Extrapolating the SFHs to $z=17$, only 11 of the 28 galaxies had already reached 20\% of their stellar mass by then. 
We derive the survey area required to detect such sources as a function of UV magnitude, finding results consistent with or slightly more optimistic than previous constraints at the bright end, and requiring larger areas than some literature estimates at the faint end. 
At $z=20$, only one galaxy of our sample already formed 20\% of its stellar mass, yielding a tentative constraint of one object at $M_{\rm UV}=-18.1\pm0.6$ per $\sim200$\,arcmin$^2$. 
Our results suggest a relatively slow decline in star formation activity between $z\sim15$ and $z\sim17$, followed by a more rapid decline between $z\sim17$ and $z\sim20$, providing new constraints on the onset of star formation in the earliest galaxies.
}
   \keywords{Galaxies: evolution, Galaxies: high-redshift }

   \authorrunning{Ciesla et al.}
   \titlerunning{Bursty SFH and redshift frontier}

   \maketitle

\section{\label{intro}Introduction}

The \textit{James Webb Space Telescope} (JWST) has dramatically advanced the identification and characterization of galaxies within the first few hundred million years after the Big Bang, enabling the detection of galaxy populations well into the epoch of reionization and approaching the era of first light. 
Early JWST surveys have revealed a substantial number of candidate galaxies at redshifts $z \gtrsim 8$ and, in some cases, spectroscopic confirmations extending to $z \sim 13$--14 \citep{CurtisLake23,Robertson23,Donnan24,Finkelstein24,Carniani24,Castellano25}. 
Pushing beyond the current spectroscopic record held by MoM-z14 at $z=14.4$ \citep{Naidu26}, several candidates at even higher redshift have been proposed on the basis of photometric methods, such as long-wavelength JWST filter dropouts or photometric-redshift estimates \citep{Castellano25,PerezGonzalez25,Kokorev25}. 
However, the reliability of these candidates remains debated and some studies report no credible candidates beyond $z\sim15$ \citep{Whitler25,McLeod26,Kreilgaard26,Weibel26}. 
Cautiously, \cite{Castellano25} analyzed the case where they assume all their $z>17$ candidates would be interlopers and discuss the impact on their results.
Such interlopers are typically either dusty galaxies at $z\sim4$--6, as in the case of a candidate initially proposed at $z=16.4$ \citep{Donnan23} and later shown to be a dusty galaxy at $z=5.1$ \citep{Zavala23}, or brown dwarf stars, as illustrated by Capotauro, a candidate $z\sim32$ galaxy \citep{Gandolfi26} subsequently identified as a brown dwarf \citep{Liu26}.
Other examples of high-$z$ galaxy candidates later confirmed as brown dwarfs are discussed in \cite{Bradac26}.
The lack of spectroscopically confirmed galaxies beyond $z=15$, in the first 270\,Myr of the Universe, raises the question whether we are hitting a physical or an observational barrier.

The uncertain nature of the highest-redshift candidates feeds directly into an ongoing debate about the evolution of the UV luminosity function (UVLF) at $z\gtrsim15$. 
Several studies, relying on the non-detection of robust candidates, favor a sharp decline of the UVLF at $z\sim15$--16 such as \cite{McLeod26} who find evidence for such a steep drop using compilation of extragalactic fields. 
This result was echoed by \citet{Weibel26}, who similarly report a rapid build-up of the observed galaxy population only from $z\gtrsim15$ onward, consistent with a late onset of active star formation. 
In parallel, using the recent cosmological parameters constrained by the DESI results \citep{Desicollab24}, \cite{Menci24} also find evidence for a steep decline of the abundance of bright galaxies in the redshift range $15\lesssim z \lesssim16$.
In contrast, studies that do report candidates beyond $z\sim15$ tend to favor a shallower decline \citep[e.g.,][]{PerezGonzalez25,Castellano25,Kokorev25}. \cite{PerezGonzalez25} find that the number density of galaxies at $z=17$ in the range $-19<M_{\rm UV}<-18$ is only a factor of $\sim4$ lower than at $z=12$, a more modest evolution than implied by a sharp cutoff.
This tension between a rapid drop of the UVLF and a more gradual decline is discussed in detail in \cite{Fontana26}.

In the absence of spectroscopically confirmed detection at $z>15$ that could be due to either observational limitations or physical constraints, a possible approach to tackle this problem would be in the reconstruction of the star formation history (SFH) of high-$z$ galaxies which would allow us to estimate or constrain when they were active and if we could observe them.
Several studies have focused on characterizing the SFH of early galaxies using different tracers, either directly from observations (e.g., the H$\alpha$-to-UV luminosity ratio, or ratios of SFR indicators sensitive to different timescales) or through SED modeling \citep[e.g.,][] {Looser25,Langeroodi24a,Ciesla24,Endsley25,Kokorev25,Carvajal25}.
There is now a consensus that at these high redshifts ($z>6$), galaxies' SFHs are dominated by bursty star-formation activity \citep[e.g.,][]{Endsley23,Dressler24,Looser24,Ciesla24,Cole25,Carvajal25,Kokorev25,Simmonds25,Clarke26,RobertsBorsani26,Chen26}.
Considerable effort has therefore gone into improving how SFHs are implemented in SED-modeling codes, in order to obtain more reliable estimates of galaxies' physical properties, from simple analytical forms with well-known biases \citep[e.g.,][]{Buat15,Ciesla17,Carnall18}, to non-analytical or non-parametric models \citep[e.g.,][]{Iyer17,Iyer19,Leja19a,Lower20,Suess22,Ciesla23}. 
More recently, to better capture the burstiness observed in early galaxies, stochastic models have been proposed, either drawn from power spectrum densities \citep[e.g.,][]{CaplarTacchella19,Tacchella20,Faisst24,Carvajal25} or built from Gaussian processes \citep{Iyer24}.

In this work, we propose, test, and apply a new SFH model designed to reconstruct, as faithfully as possible, the SFHs of galaxies at $z\geq8$.
The distinctive feature of this model is that it allows the strength of the stochastic bursts to be modulated with time, enabling a transition between a bursty phase and a smooth one at any point in the galaxy's history. 
The paper is organized as follows. In Sect.~\ref{sfh} we present this new formalism and calibrate it against \textsc{sphinx$^{20}$} mock catalogs and their corresponding SFHs \citep{Katz23}. In Sect.~\ref{evolpath}, we apply this model to the UNICORN/GOODS-South sample (Finkelstein et al., in prep.) and recover the SFHs of galaxies at $z\geq9$. 
We use these reconstructed SFHs to discuss the formation epoch of these galaxies, and place our results in the context of the dearth of galaxies at $z\geq15$ in Sect.~\ref{dearth}.
We then provide tentative constraints on star formation at $z\sim17$ in Sect.~\ref{backintime}.
Throughout the paper, we use a \cite{Chabrier03} initial mass function and WMAP7 cosmology \citep{Komatsu11}.

\section{\label{sfh}Bursty star formation history for SED modeling}

Bursty star formation histories are now routinely used to model the star formation histories of galaxies in SED modeling codes.
Current models use SFH drawn from power spectrum density \citep[e.g.,][]{CaplarTacchella19,Faisst24,Carvajal25} or from Gaussian process \citep[e.g.,][]{Iyer24}.
In this work, we propose a new approach based on an Ornstein-Uhlenbeck process, or mean reverting process, which is a continuous-time stochastic process modeling random fluctuations with a tendency to drift back toward a long-term mean over time. 
In the following section, we present how we define and implement this SFH in \textsc{cigale} \citep{Boquien19} and test this model using simulations.

\subsection{\label{variance} Implementation}

\begin{figure*}[h!]
    \includegraphics[width=\textwidth]{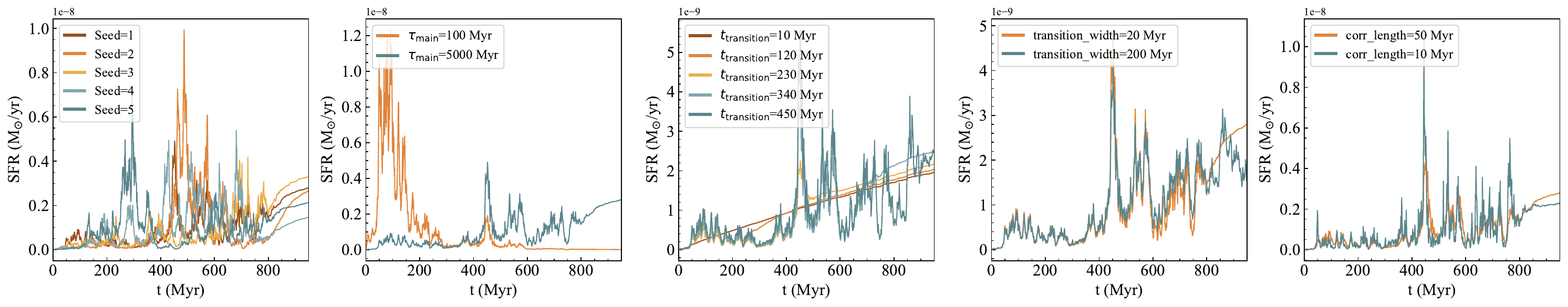}
    \caption{
    Examples of SFHs generated with the \textsc{sfhbursty} module, illustrating the effect of the main input parameters on the resulting SFH shapes. In each panel, only one parameter is varied, all others are fixed. From left to right, the panels show the effect of the random seed, the delayed-\(\tau\) timescale \(\tau_{\rm main}\), the time of transition between the two burstiness levels, the duration of that transition, and the correlation time of the stochastic kicks. The SFHs are normalized to 1\,M$_{\odot}$.
    }
    \label{example_sfh}
\end{figure*}

The \textsc{sfhbursty} module generates SFHs by starting from a smooth delayed-\(\tau\) component and multiplying it by a correlated stochastic fluctuation. The model is designed to separate three ingredients with simple physical interpretations: \(\tau_{\rm main}\) controls the smooth, long-term assembly history; \(t_{\rm corr}\) controls how long bursty deviations remain coherent; and \(\sigma_{\rm kick}(t)\) controls the strength of the burstiness as a function of time.

The smooth component is a delayed-\(\tau\) SFH,
\[
\mathrm{SFR}_{\rm smooth}(t)
=
\frac{t}{\lvert\tau_{\rm main}\rvert}
\exp\left(-\frac{t}{\tau_{\rm main}}\right),
\]
where \(t\) is sampled on a uniform 1 Myr grid. This component describes the underlying, slowly varying SFH before short-timescale stochasticity is added.

We then introduce bursty fluctuations in logarithmic SFR space. This means that the stochastic component acts multiplicatively on the SFR, rather than additively. Let \(x_i\) be the log-SFR fluctuation at timestep \(i\). We model it as a correlated process,
\[
x_i = a x_{i-1} + \eta_i,
\]
with
\[
a = \exp\left(-\frac{\Delta t}{t_{\rm corr}}\right),
\]
where \(\Delta t = 1\,\mathrm{Myr}\). The parameter \(t_{\rm corr}\) sets the coherence time of the fluctuations. If \(t_{\rm corr}\) is small, the stochastic component changes rapidly from one timestep to the next. If \(t_{\rm corr}\) is large, deviations from the smooth SFH persist for longer.
The term \(\eta_i\) is a stochastic kick. It is drawn from a Gaussian distribution,
\[
\eta_i \sim \mathcal{N}\left(0,\sigma_{\rm kick}^2(t_i)\right).
\]
The quantity \(\sigma_{\rm kick}(t)\) sets the amplitude of the kicks, and therefore the burstiness of the SFH. We allow this amplitude to evolve smoothly with time using a sigmoid form,
\[
\sigma_{\rm kick}(t)
=
\sigma_{\rm floor}
+
(\sigma_0-\sigma_{\rm floor})
\left[
1+\exp\left(\frac{t-t_{\rm transition}}{w}\right)
\right]^{-1}.
\]
Here, \(\sigma_0\) and \(\sigma_{\rm floor}\) set the two asymptotic burstiness amplitudes, \(t_{\rm transition}\) sets when the transition occurs, and \(w\) controls how gradual the transition is. This form allows the SFH to be more bursty at early times and smoother at late times, or the reverse, depending on the relative values of \(\sigma_0\) and \(\sigma_{\rm floor}\).

The final SFH is
\[
\mathrm{SFR}(t_i)
=
\mathrm{SFR}_{\rm smooth}(t_i)
\exp\left[
x_i-\frac{1}{2}V_i
\right],
\]
where \(V_i = \mathrm{Var}(x_i)\) is the variance of the log-SFR fluctuation at timestep \(i\). The subtraction of \(V_i/2\) is the standard lognormal correction. It ensures that the stochastic multiplicative factor has mean unity, so that \(\mathrm{SFR}_{\rm smooth}(t)\) remains the mean SFH rather than being biased upward by the exponential transformation.

Because \(\sigma_{\rm kick}(t)\) can change with time, the stochastic component is not necessarily stationary. However, its temporal memory is still controlled by \(t_{\rm corr}\). The stochastic kicks are smoothed into persistent excursions whose typical duration is set by this correlation timescale.

This prescription is intentionally phenomenological. The stochastic kicks are not assumed to correspond to a single physical process. They represent the net effect of mechanisms capable of producing short-timescale departures from the smooth SFH, such as bursty gas accretion, feedback-regulated gas cycling, interactions, or unresolved star-forming regions. The resulting SFH therefore consists of a smooth delayed-\(\tau\) component modulated by temporally correlated stochastic fluctuations whose amplitude can evolve with time.

Concretely, in the \textsc{sfhbursty} module, the input parameters are:
\begin{itemize}
    \item \textsc{age}: time since the onset of star formation, in Myr.
    \item \textsc{n\_sfh}: number of random realizations generated for each parameter set. Equivalently, this controls the number of random seeds used to sample the stochastic component.
    \item \textsc{tau\_main} ($\tau_{\rm main}$): e-folding time of the underlying delayed-\(\tau\) SFH, in Myr. This parameter controls the smooth, long-term assembly history.
    \item \textsc{sigma0} ($\sigma_0$): initial amplitude of the stochastic kicks. Larger values produce more bursty SFHs; values larger than \(0.6\) typically lead to highly bursty histories \citep[see][]{Carvajal25}.
    \item \textsc{sigma\_floor} ($\sigma_{\rm floor}$): final amplitude of the stochastic kicks. Lower values produce smoother SFHs; values below \(0.1\) lead to histories close to the underlying delayed-\(\tau\) SFH \citep[see][]{Carvajal25}.
    \item \textsc{t\_transition} ($t_{\rm transition}$): time at which the transition between the two burstiness amplitudes occurs, in Myr.
    \item \textsc{transition\_width} (\(w\)): duration of the transition between \textsc{sigma0} and \textsc{sigma\_floor}, in Myr. Small values produce a sharp transition, while larger values produce a smoother evolution of the burstiness amplitude.
    \item \textsc{corr\_length} ($t_{\rm corr}$): correlation time of the stochastic kicks, in Myr. Small values produce rapidly varying, short-lived fluctuations, while large values produce coherent deviations from the smooth SFH over longer timescales.
\end{itemize}

Examples of the impact of each of these parameters in the produced SFHs are shown in Fig.~\ref{example_sfh}. The module is publicly available\footnote{\url{https://gitlab.lam.fr/cigale/cigale/-/commit/421b9d47c7be97d5384350bcd0cf204b553b27c6}} and can be used with the \textsc{cigale} version 2025.1 and later.

\subsection{\label{calib} Test and calibration on the \textsc{sphinx$^{20}$} simulations}

\begin{figure}[h!]
    \includegraphics[width=\columnwidth]{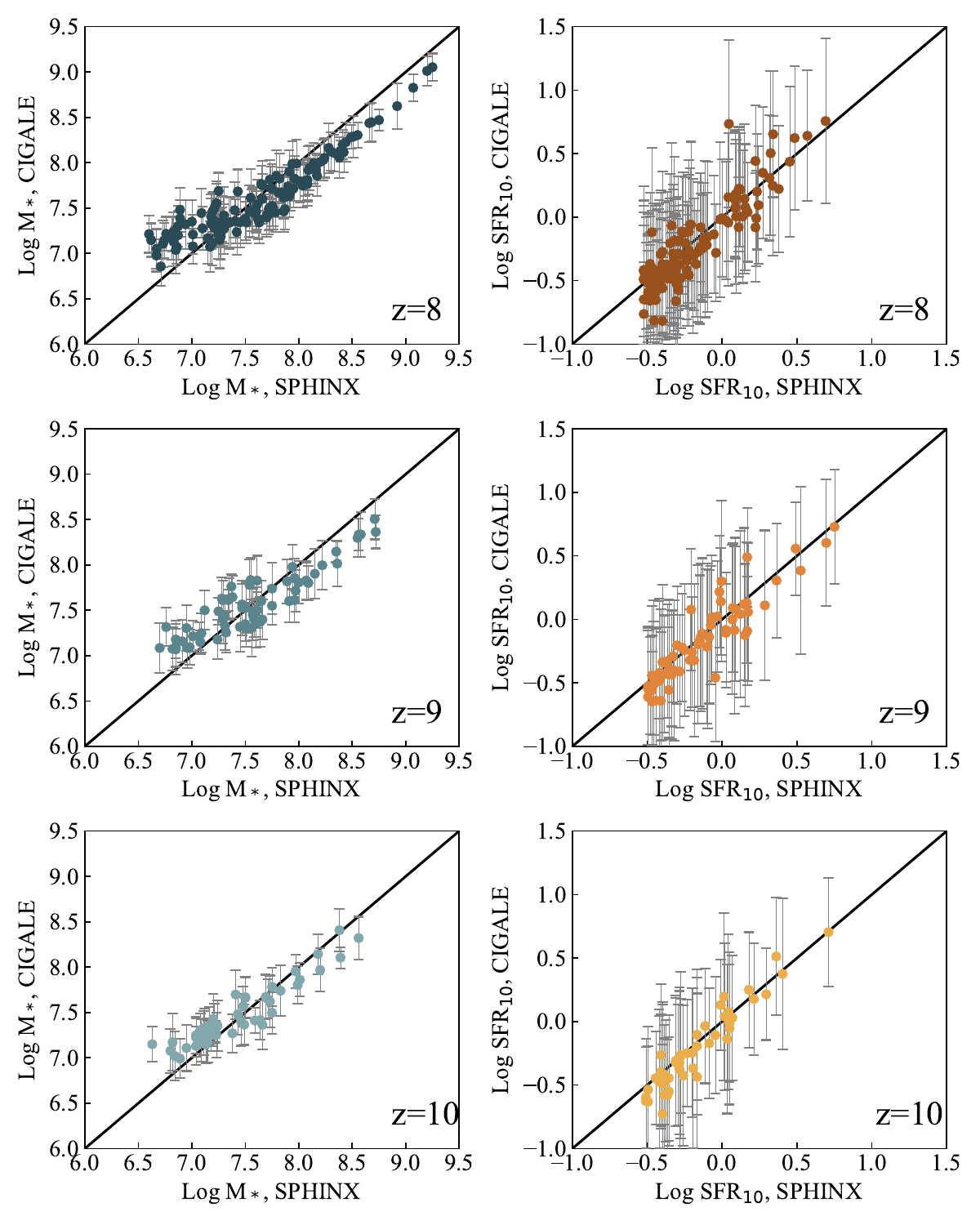}
 \caption{Comparison between the stellar masses and SFRs of the $z=8$ (top), $z=9$ (middle), and $z=10$ (bottom) \textsc{sphinx} galaxies and those obtained from the associated SEDs fitted with the stochastic SFH presented in this work. SFR from both \textsc{sphinx$^{20}$} and \textsc{cigale} are averaged over 10\,Myr. The bursty SFH recovers relatively well the stellar masses and SFR of the \textsc{sphinx$^{20}$} galaxies at $z\geq8$.}
 \label{comp_mstar_sfr}
\end{figure}

We test the suitability and ability of this new SFH model in recovering basic properties of high-$z$ galaxies (stellar masses and SFRs) and reconstructing the SFH of these galaxies.
To do so, we make use of the \textsc{sphinx$^{20}$} set of simulations \citep{Katz23}, which includes photometric catalogs computed on JWST filter sets.
We run \textsc{cigale} on their mock photometric catalog of $z=8, 9, 10$ galaxies, using the \textsc{sfhbursty} module.
Input parameters used for this are provided in Table~\ref{input_param}.
This parameter set results from calibration tests consisting of systematically varying each of them, aiming at providing the best agreement with the \textsc{sphinx$^{20}$} simulations.
This led to the unusual negative value of $\tau_{\rm main}$ which provides a slow increase followed by a more rapid one of the SFH.
A more typical value yields unsatisfactory results for the tests described below.
The parameters $t_{\rm transition}$, transition\_width, and corr\_length, which handle the transition between stochastic and smooth SF activities, are fixed as their variation does not impact the quality of the comparison with the \textsc{sphinx$^{20}$} simulations.
Furthermore, we use only the dust-free SEDs as, even if there are examples of galaxies with significant dust content \citep{Rodighiero26}, at population level the attenuation by dust is found to be very low at these redshifts \citep[e.g.,][]{Ziparo23,Ciesla25,Bakx26}. 

\begin{table}[!htbp]
   \centering
   \caption{\textsc{cigale} input parameters used to fit \textsc{sphinx$^{20}$} $z=8, 9, 10$ mock SEDs, that are different from the default value. This set of parameters is also used on real data.}
   \label{input_param}
   \begin{tabular}{l c l}
   \hline\hline
   Parameter & Value & Definition\\
   \hline
   \multicolumn{3}{c}{\textsc{sfhbursty}}\\
   \hline
   $age$ (Myr) & 10 values between & age of first star\\
               & 50 and 630\,Myr & \\
   $N_{\rm SFH}$ & 7000 & \# of SFH seeds \\
   $\tau_{\rm main}$ (Myr) & $-100$\tablefootmark{a} & e-folding time \\
   $\sigma_0$ & 0.1, 1.0 & initial kick amplitude\\
   $\sigma_{\rm floor}$ & 0.1, 1.0 & final kick amplitude\\
   $t_{\rm transition}$ (Myr) & 30, 100, 200, 300 &\\
   transition\_width (Myr) & 20 &\\
   corr\_length (Myr) & 5 &\\
   \hline
   \multicolumn{3}{c}{Stellar population -- \textsc{bc03}}\\
   \hline
   metallicity & 0.0004, 0.004 & \\
   \hline
   \multicolumn{3}{c}{Emission lines -- \textsc{nebular}}\\
   \hline
   $\log$U & $-2$ & ionization parameter\\
   \hline
   \multicolumn{3}{c}{Dust attenuation -- \textsc{dustatt\_modified\_starburst}}\\
   \hline
   E(B-V) lines & 0.0 & \\
   \hline
   \end{tabular}
   \tablefoot{
   \tablefoottext{a}{This negative value allows for a rapid increase of the SFR.}
   }
\end{table}

In Fig.~\ref{comp_mstar_sfr}, we compare the stellar masses obtained with \textsc{cigale} to the true ones provided by \textsc{sphinx$^{20}$}.
The use of the \textsc{sfhbursty} module provides a good estimate of M$_\star$ with a scatter of 0.25, 0.24, and 0.18\,dex in the relation at $z=8$, $z=9$, and $z=10$, respectively.
We note a systematic offset from the one-to-one relationship, especially at high masses, which we attribute to the use of BPASS stellar population models \citep{Eldridge16} by \textsc{sphinx$^{20}$} while in \textsc{cigale} we assume \cite{BruzualCharlot03} models \citep[see also][]{Bozhidara26}.
The SFR, here averaged over the last 10\,Myr, is also well recovered by the \textsc{sfhbursty} SFH module, with no specific systematic and a scatter of 0.15, 0.13, and 0.10\,dex at $z=8$, $z=9$, and $z=10$, respectively.  

\begin{figure*}[h!]
    \includegraphics[width=6cm]{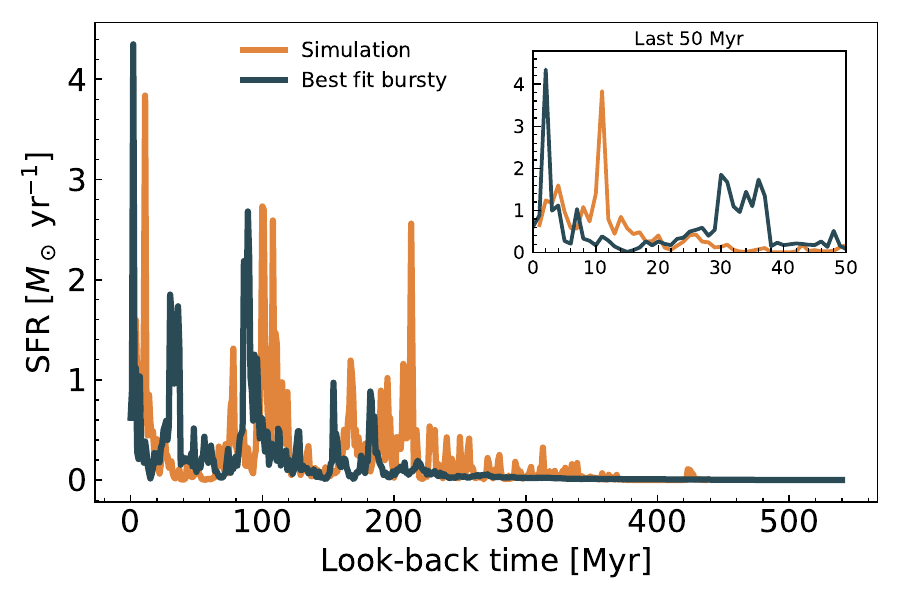}
    \includegraphics[width=6cm]{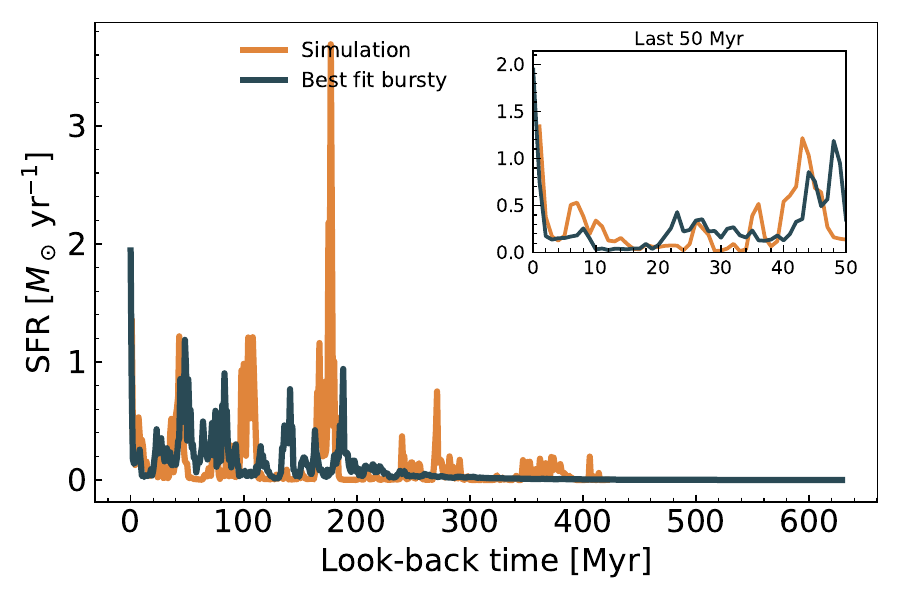}
    \includegraphics[width=6cm]{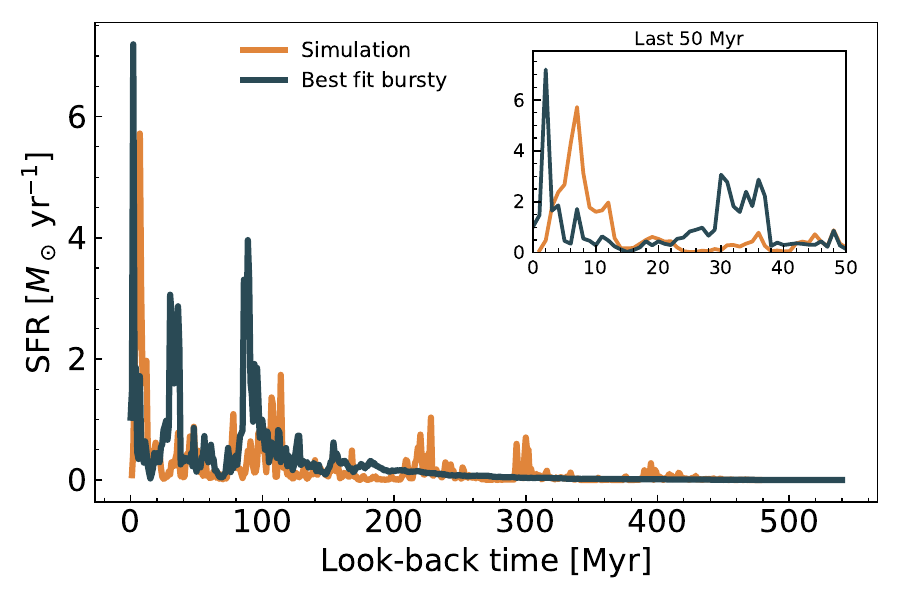}
 \caption{Examples of reconstructed SFH using the \textsc{sfhbursty} SFH module. In each panel, the true SFH from \textsc{sphinx$^{20}$} is shown in orange, while the reconstructed one is dark blue. Although the bursty SFH does not reproduce every single burst of the true SFH, it provides a close estimate of the predicted \textsc{sphinx$^{20}$} SFHs. There is a good agreement between the true and the estimated SFHs in the last 50\,Myr.}
 \label{reconst_sfh}
\end{figure*}

\begin{figure*}[h!]
    \includegraphics[width=6cm]{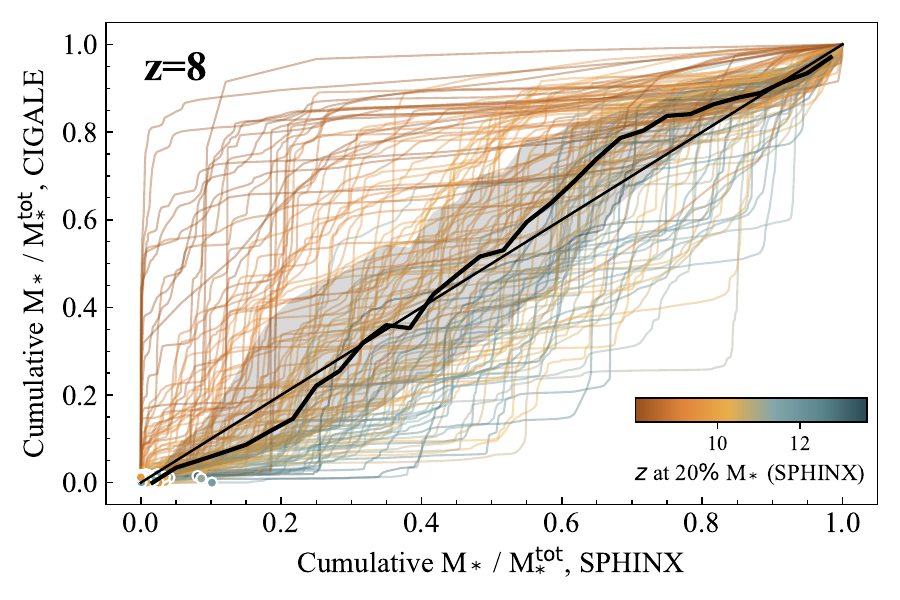}
    \includegraphics[width=6cm]{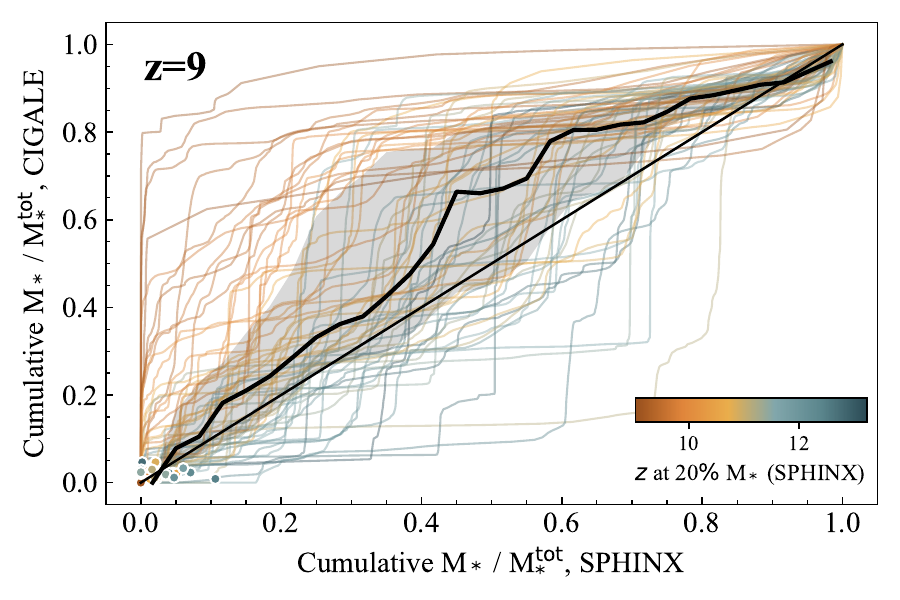}
    \includegraphics[width=6cm]{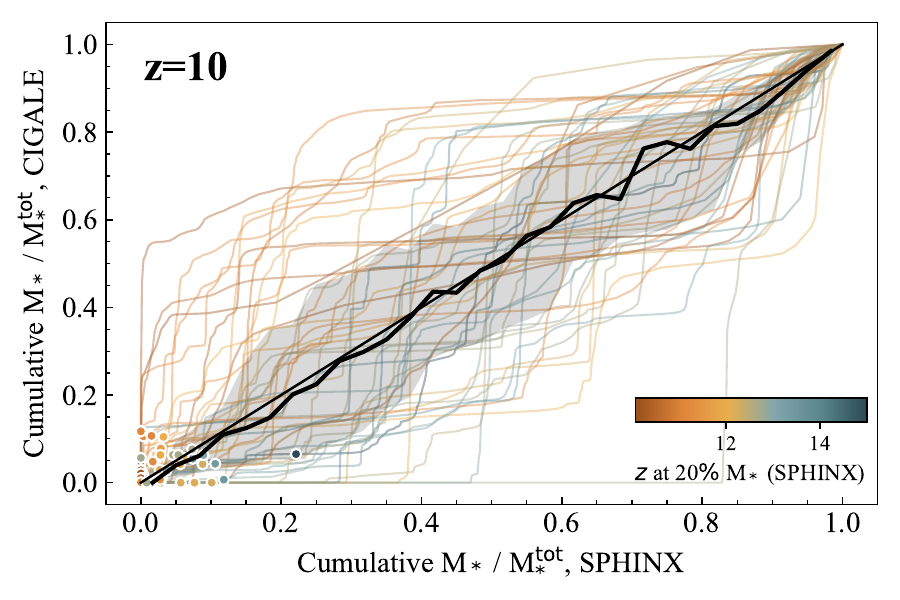}
 \caption{Reconstructed cumulative mass history using the \textsc{sfhbursty} SFH module. In each panel, the true cumulative mass from \textsc{sphinx$^{20}$} is on the x-axis, while the reconstructed one on the y-axis. The solid black line shows the median cumulative history recovered by the \textsc{sfhbursty} SFH module and the grey shaded region delimits the 25$^{th}$ to 75$^{th}$ percentile. A perfect reconstruction would yield a one-to-one relation between the recovered and the true cumulative masses with no dispersion. The bursty SFH allows, at population level, to recover the SFH of early galaxies as shown through this comparison of their cumulative SFH.}
 \label{cumulatives}
\end{figure*}

To test the ability of the \textsc{sfhbursty} module in reconstructing the overall SFH of the \textsc{sphinx$^{20}$} simulated galaxies, we show in Fig.~\ref{reconst_sfh}, three examples comparing the true SFH from \textsc{sphinx$^{20}$} and the reconstructed one using the \textsc{sfhbursty} SFH.
There is a striking agreement between the two in the last $\sim$50\,Myr, and an overall good agreement on longer timescales.
We stress here that the goal of the \textsc{sfhbursty} SFH is not to recover every single burst of the true SFH, but rather to propose a model that is close to what we expect the SFH of early galaxies to be.
We investigate the recovered SFH looking at the build-up of stellar mass from the cumulative mass histories in Fig.~\ref{cumulatives}.
For each time step (1\,Myr), we calculate the mass formed up to this particular time and the corresponding fraction of the total stellar mass.
We then compare the obtained cumulative mass from the SFHs of the \textsc{sphinx$^{20}$} simulation and from the reconstructed ones using the \textsc{sfhbursty} SFH module.
The median cumulative mass histories obtained with \textsc{cigale} at $z=8$, $z=9$, and $z=10$ are close to the one-to-one relation, especially at $z=8$ and $z=10$.
The dispersion around this median cumulative is relatively narrow indicating that, even though the cumulative of some individual galaxies are not well recovered, at population level the \textsc{sfhbursty} SFH is able to reproduce the bursty SFHs of early galaxies.

\begin{figure*}[h!]
    \includegraphics[width=6cm]{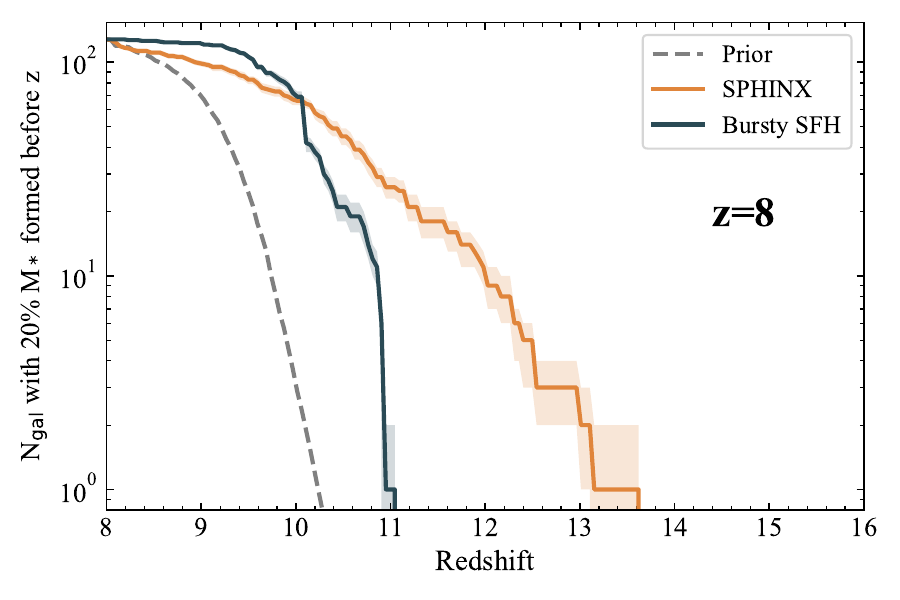}
    \includegraphics[width=6cm]{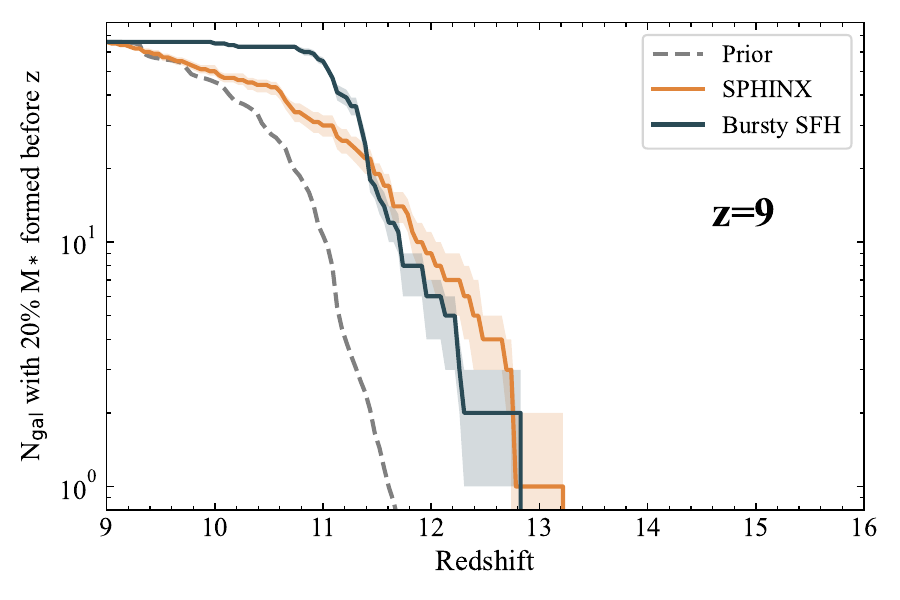}
    \includegraphics[width=6cm]{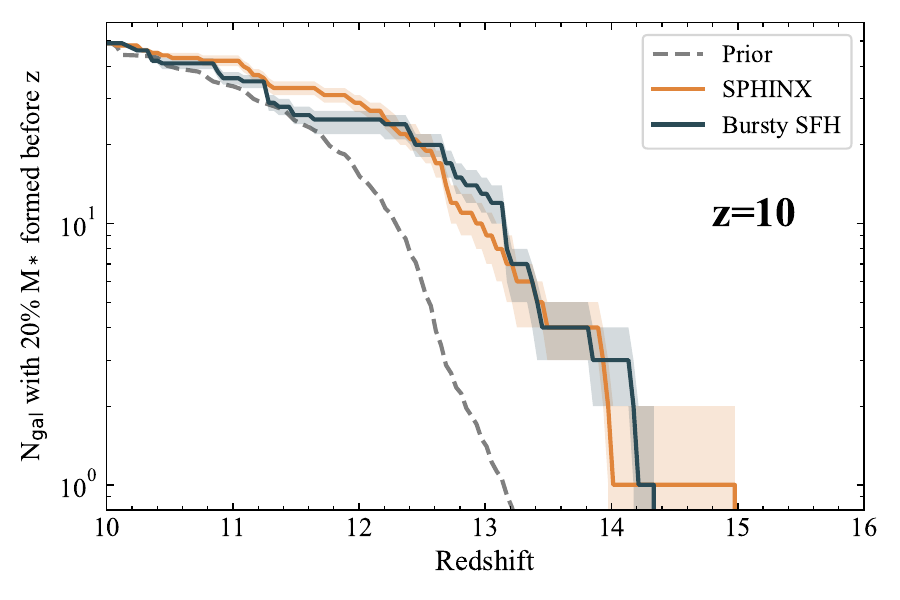}
 \caption{Cumulative number of galaxies from \textsc{sphinx$^{20}$} observed at $z=8$ (left), $z=9$ (center), and $z=10$ (right) that had formed 20\% of their final stellar mass before a given redshift $z$. The orange curve uses the \textsc{sphinx$^{20}$} SFH. The dark blue curve uses the SFHs reconstructed with the \textsc{sfhbursty} module. Shaded regions show the 25$^{th}$–75$^{th}$ percentile range from bootstrap resamplings of the galaxy sample. The grey dashed lines indicate the distribution from the prior.}
 \label{ngalbeforez}
\end{figure*}

To test how well the formation epoch of the galaxies can be recovered by the \textsc{sfhbursty} model, we compare the true redshift at which 20\% of the stellar mass has been reached by the \textsc{sphinx$^{20}$} galaxies with the value obtained from the reconstructed SFH. 
Figure~\ref{ngalbeforez} shows the resulting cumulative distributions at $z=8$, $9$ and $10$. 
At $z=8$, \textsc{cigale} recovers the typical formation epoch well: the recovered median $z_{\rm 20}$ is 10.06, against a true value of 10.11, and there is no median offset. 
However, the shape of the distribution is not reproduced. 
The recovered values span a much narrower range than the true ones. 
The 25–75\% interval is [9.48, 10.50] while the true one is [8.68, 11.25] and never exceeds $z\sim11$, whereas the true $z_{\rm 20}$ reaches 13.6. 
The largest difference between the curves is 28 galaxies at $z\sim10.4$. 
The good agreement in the median therefore results from individual errors that cancel out. 
Early-forming galaxies are assigned formation epochs that are too late, and late-forming ones epochs that are too early. 
This behavior is expected when the early part of the SFH is weakly constrained by the SED.
In addition, the upper end of the recovered distribution is set by the prior, which contains no model with $z_{\rm form,20\%}\gtrsim11$ at this redshift.
Extending the prior to earlier formation epochs produced a large number of spurious galaxies with $z_{\rm form,20\%}>15$, showing that at $z=8$ the early SFH is not constrained by the photometry alone.
At $z=9$, this compression comes with a systematic bias. 
The recovered $z_{\rm 20}$ is too early for 68\% of the galaxies. 
The recovered distribution is again narrower than the true one but for the oldest galaxies of the simulations, with $z_{\rm 20}\gtrsim11.5$, the agreement between \textsc{sphinx$^{20}$} and the reconstructed SFH is good. 
In contrast, at $z=10$, the recovered distribution is statistically consistent with the true one (KS p = 0.54) and has a similar width ([11.22, 12.71] against [10.91, 13.05]). 
There is no significant systematic shift, and the largest difference between the curves drops to 11 galaxies at $z\sim11.7$. 
The \textsc{sfhbursty} model misses the one \textsc{sphinx$^{20}$} galaxy that has $z_{\rm 20}=15$ placing it at 14.3.
Overall, the \textsc{sfhbursty} model recovers the population-averaged epoch of early assembly, but $z_{\rm 20}$ for individual galaxies is uncertain by about one unit in redshift. 
From $z\geq9$, the recovery of $z_{\rm form,20\%}$, or equivalently age$_{\rm 20}$, is good for the oldest galaxies.

\section{\label{evolpath} Evolutionary pathways of early galaxies}

Now that we have tested this new way of modeling galaxies' SFHs, we apply this model to observations with the goal of reconstructing the SFH of galaxies at $z>9$.

\subsection{The UNICORN/GOODS-South sample}

We use the catalog obtained as part of the UNICORN (Uniform Near-Infrared CatalOg from Robust imagiNg) project, which provides homogeneous data reduction and source extraction for all JWST extragalactic fields (Finkelstein et al., in prep.), including the GOODS-South field observed as part of the JADES survey \citep[Data Release 5,][]{Johnson26}.
We briefly describe the source extraction and photometry process below, and refer the reader to \cite{Finkelstein24} for further details.
Images from JADES Data Release 5 are used to produce the GOODS-South galaxy catalog over a region of 199\,arcmin$^2$.
The point spread functions (PSFs) of all ACS bands and all NIRCam bands bluer than F277W are matched to the F277W PSF.
For bands with larger PSFs (the redder NIRCam bands and all WFC3 bands), a correction is calculated and applied by convolving the F277W image to the larger PSF and measuring the flux ratio between the native and convolved images.
The 5$\sigma$ detection limits are 29.85 and 29.47\,mag AB in the F277W and F444W filters, respectively.
Photometry was performed with Source Extractor \citep[v2.25.0;][]{Bertin96} in dual-image mode, using a weighted sum of the native-resolution F277W and F356W images as the detection image.
Flux uncertainties were derived by placing random, non-overlapping apertures with diameters of 0.1–1.5\arcsec in empty regions of the image (avoiding the segmentation map), and fitting a four-parameter function to the normalized median absolute deviation in each aperture as a function of aperture area.
The flux error for each object is then calculated from this fit, evaluated at the area of that object's aperture.
Fiducial fluxes are measured in small Kron apertures, to which aperture corrections are applied.
In the NIRCam bands, we ran source injection simulations, injecting small Sérsic profiles generated with Galfit into the F277W and F356W images.
We then created a detection image from F277W+F356W and measured photometry in F277W.
Comparing input and output fluxes, we find that the input fluxes are ~5–20\% higher than the output fluxes, with the discrepancy increasing slightly toward fainter magnitudes.
Fluxes are corrected for this offset, by a factor of typically 1.02 at m=24\,mag, rising to 1.10 at m=28\,mag, and capped at 1.20 for the faintest magnitudes.
The resulting v0.91 catalog comprises 22 photometric bands, including four HST/ACS filters from CANDELS \citep{Grogin11,Koekemoer11}, eight JWST/NIRCam wide filters, and ten JWST/NIRCam medium bands.
Photometric redshifts were estimated with EAzY \citep{Brammer08}, following the methodology of CEERS Key Paper I \citep{Finkelstein23}, including new templates from \cite{Larson22}.
In this work, we select galaxies with $9<z_{\rm phot}<15$ that are flagged as satisfying all quality criteria defined by the UNICORN project.
We list the main criteria below:
\begin{itemize}
    \item The photometric redshift is $9<z_{\rm phot}<15$.
    \item The signal-to-noise ratio (S/N) in at least two of the aperture fluxes measured across all NIRCam bands must exceed 5.3, or three of these fluxes must have S/N greater than 4.3.
    \item The value of the error map at the central pixel of a given object must be $<10^3$ in the F115W, F150W, F277W, and F444W filters.
    \item The integral of P($z$) is $\geq$0.7.
    \item The $\chi^2$ of the best-fitting photometric redshift is $\leq$60.
    \item No filters lying fully blueward of the Ly$\alpha$ break have S/N$>$2.5, and no more than one has S/N$>$2. The break wavelength is set at the redshift corresponding to the 16\% of the full P($z$).
    \item A galaxy is retained if $z_{\rm phot}>9$ and the secondary low-redshift solution, defined as the peak of P($z$) at $z<7$, has a $\chi^2$ at least 4.0 higher than $\chi^2_{\rm min}$.
\end{itemize}

This selection results in a sample of 400 galaxies with $9<z<15$ with a median M$_{\rm UV}$ of $-18.20^{+0.54}_{-0.78}$.

\subsection{\label{sec:sfh} Star formation histories of $9<z<15$ galaxies}

\begin{figure}[h!]
    \includegraphics[width=\columnwidth]{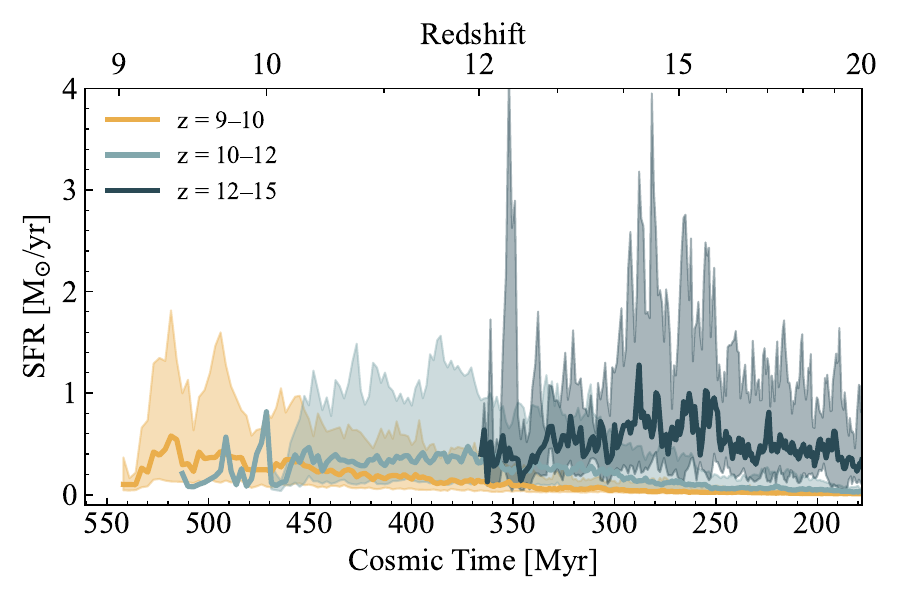}
 \caption{Median SFH of our galaxies, binned by redshift (colored solid lines). The shaded regions show the 25$^{\rm th}$ and 75$^{\rm th}$ percentile. Galaxies at $z=12-15$, detected in our survey, have formed much earlier than galaxies at $z<12$ and show an increasing SFH reaching a peak followed by a slow declining phase.}
 \label{unicorn_sfh}
\end{figure}

We fit the UNICORN/GOODS-South sample with \textsc{cigale} using the same set-up described in Sect.~\ref{calib}.
The median and dispersion of the reconstructed SFHs are shown for the whole sample in Fig.~\ref{unicorn_sfh} in different redshift bins.
They indicate a rapid and recent growth, especially for galaxies at $z \lesssim 10$.  
Galaxies at $z\gtrsim12$ show a wider variation in SFH shape and the median SFH shows a peak around $z\sim13-14$ with a slight decrease at later time.
Since we are detecting the brightest and most massive galaxies at these high redshifts, the median SFH may be biased toward these massive galaxies.

\begin{figure}[h!]
    \includegraphics[width=\columnwidth]{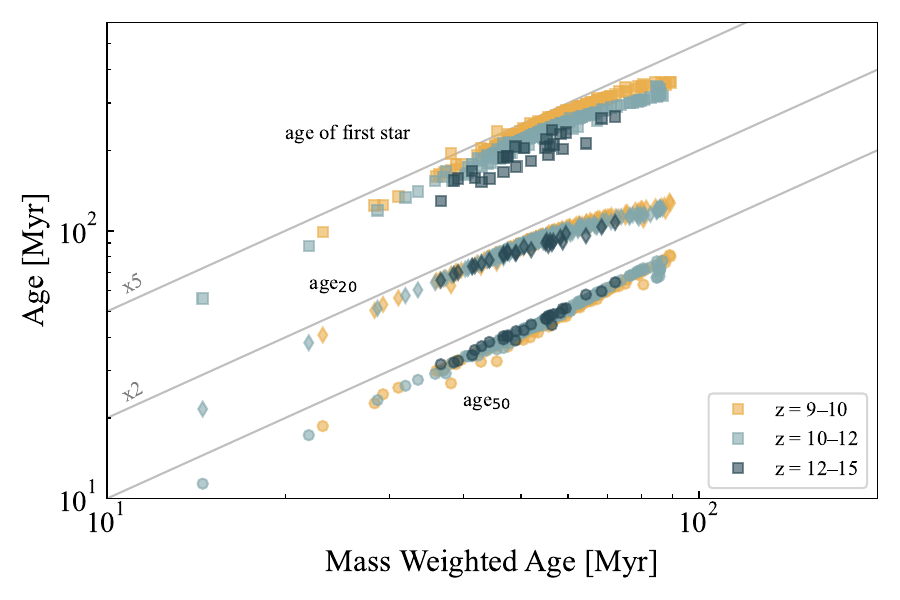}
 \caption{Different age estimates as a function of mass-weighted age for the galaxies of our sample. Squares indicate the age defined as the duration of the whole SFH, diamonds show age measured from the time when galaxies formed 20$\%$ of their stellar mass, and circles the age measured from the time when they formed 50$\%$ of their stellar mass. The symbol's color indicates the redshift bins.}
 \label{agedef}
\end{figure}

To evaluate the age of the galaxies, we consider $age_{\rm 20}$, the time elapsed since the galaxies formed 20\% of their stellar mass.
Other age definitions can be used such as $age_{\rm 50}$ when galaxies formed 50\% of their mass, the total length of the SFH obtained from the best fit, and the mass-weighted age.
In Fig.~\ref{agedef}, we compare $age_{\rm 20}$, $age_{\rm 50}$, and the age of the first star to the mass-weighted age.
To first order, $age_{\rm 50}$ is the closest estimate to a mass-weighted age, while $age_{\rm 20}$ is approximately two times the mass-weighted age, and the age of the first star five times the mass-weighted age.

\begin{figure}[h!]
    \includegraphics[width=\columnwidth]{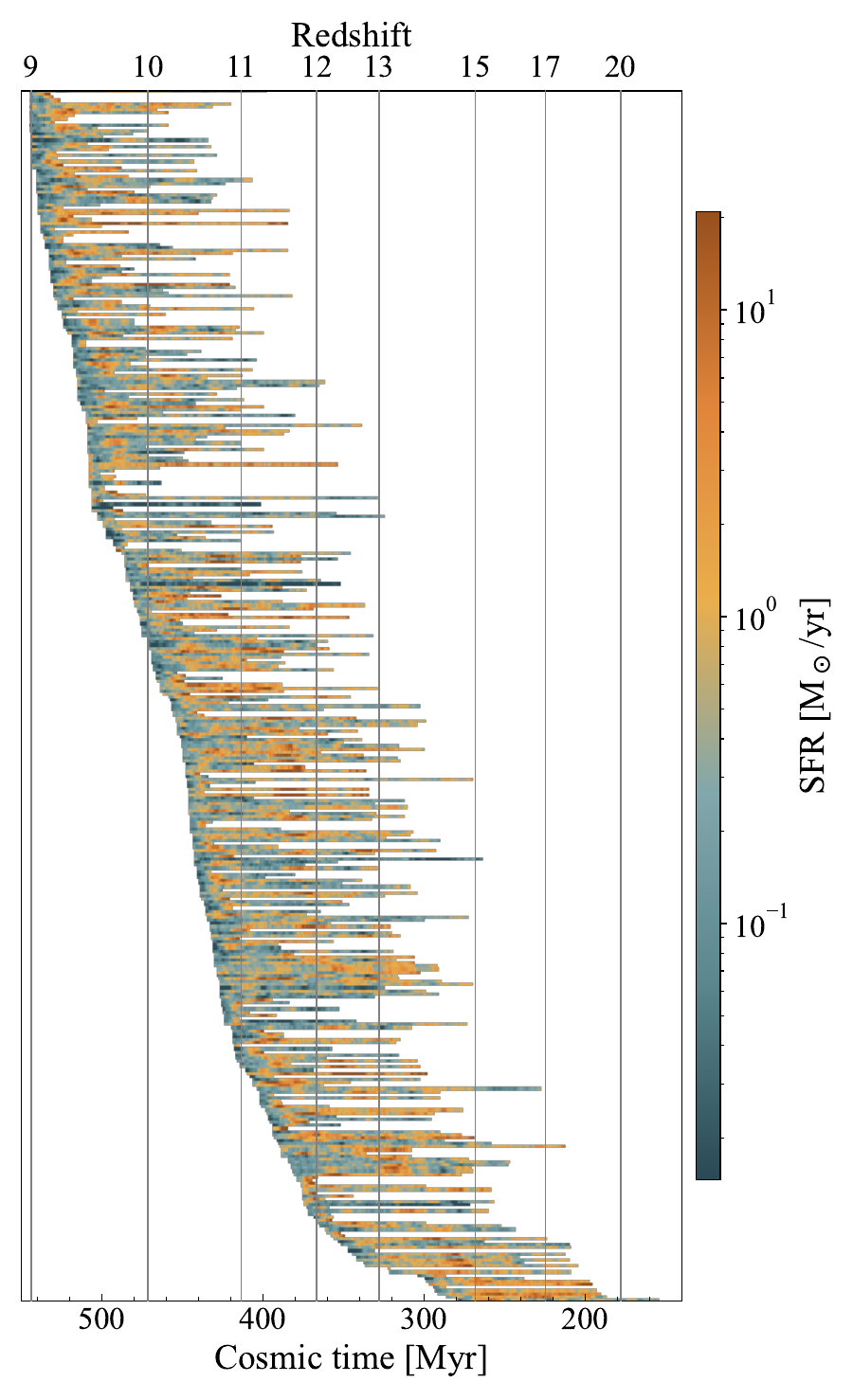}
 \caption{SFH visualization of the galaxies of our sample, ranked by observed redshift from top to bottom. Each line is the SFH of a galaxy, starting when the galaxies have reached 20\% of their stellar mass. They are color-coded by SFR, showing the burstiness of their star formation activity. Galaxies at $z<10$ are younger and did not reach 20\% of their stellar mass before $z=12$ implying that they are not the descendant of the spectroscopically confirmed sources at $z\gtrsim13$. 
 }
 \label{unicorn_ridge}
\end{figure}

\begin{figure}[h!]
    \includegraphics[width=\columnwidth]{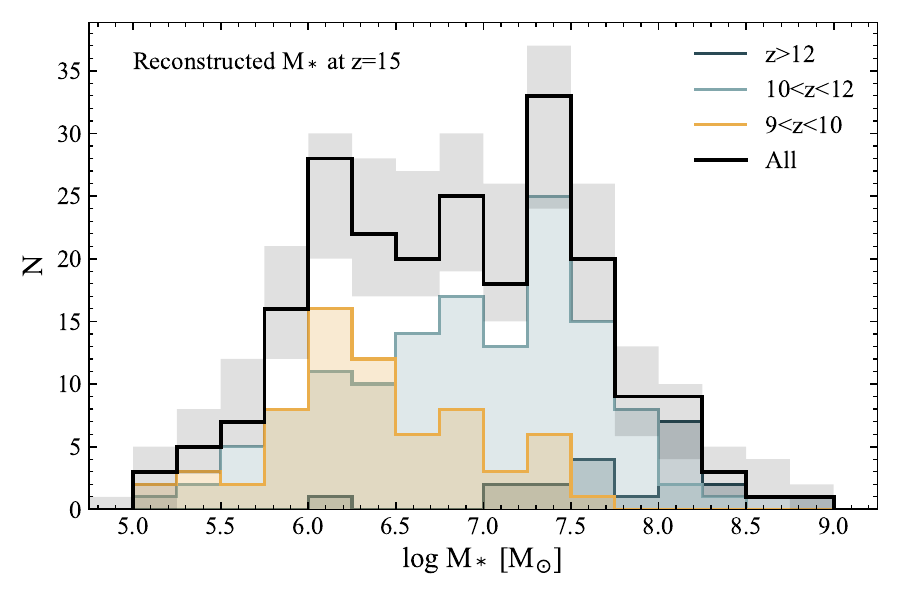}
 \caption{Distribution of the stellar masses taken at $z=15$ from the reconstructed SFH for all the galaxies of our sample. The SFHs derived with \textsc{cigale} indicate that 56\% of our galaxies had already assembled a non-zero stellar mass by $z=15$.
 }
 \label{unicorn_mz15}
\end{figure}

In Fig.~\ref{unicorn_ridge}, we show the SFH of all individual galaxies in our sample, color-coded by SFR and ranked by increasing redshift.
The color-coding highlights the burstiness of their star-formation activity, with peaks reaching SFR higher than 20\,M$_{\odot}$yr$^{-1}$, as well as periods of very low activity, consistent with no star formation.
Taking the time at which galaxies had formed 20\% of their stellar mass as the onset of their SFH, galaxies at $z<10$ formed no earlier than $z=12$, and none of them have already formed 20\% of their stellar mass by $z\geq15$.
Only 5\% of the $z=10-12$ galaxies had reached 20\% of their mass by $z=15$, whereas 70\% of the $z=12-15$ sources had.
Therefore, using this definition of age, the bulk of the population at $z\sim9$ is unlikely to be the descendants of the $z>13-14$ galaxies detected and spectroscopically confirmed \citep[e.g.,][]{Carniani24,Naidu26,Schouws25}.
However, integrating the best-fit SFHs, we find that 56\% of our sample had already formed stars by $z=15$. 
We show the distribution of the stellar mass in place at $z=15$ in Fig.~\ref{unicorn_mz15} in three different redshift bins.
The distribution spans $10^{5}$–$10^{9}\,M_\odot$, and its position depends strongly on the observed redshift. 
Galaxies observed at $z>12$ had already assembled larger stellar masses by $z=15$ than those observed at $9<z<10$, which mostly populate the low-mass end of the distribution.
An absolute criterion on the age, such as a threshold in the mass formed, rather than a percentage of the mass formed at time of observation indicates that, at $z=15$, 24\% of our sample has already formed 10$^7$\,M$_{\odot}$ (Fig.~\ref{unicorn_mz15}).
Their median redshift is $z=10.9^{+0.9}_{-0.5}$, confirming though that the bulk of the galaxies at $z\sim9$ are unlikely the descendants of the $z>13-14$ galaxies.

We now investigate the descendants of these massive galaxies at $z>13$.
The median stellar mass of the 11 sources of our sample at $z>13$ is $\log(M_\star/M_\odot)=8.2$.
Assuming that they will keep forming stars at a rate equal to their average SFR over the last 50\,Myr down to $z=9$, their median stellar mass by then will be $\log(M_\star/M_\odot)=8.8$.
In our $9<z<10$ sample, there are 2 galaxies with a stellar mass equal or larger than $\log(M_\star/M_\odot)=8.8$.
They were both forming stars at $z=15$ according to their reconstructed SFH, and had a stellar mass larger than $\log(M_\star/M_\odot)=7.3$, barely 3\% of their stellar mass at time of observation.
Therefore, the detected $z>13$ galaxies represent the tip of the mass distribution of the galaxy population at these high redshifts.
Furthermore, as seen in Fig.~\ref{unicorn_sfh}, the SFHs of these galaxies tend to slowly decline making their descendants even more difficult to observe in a UV rest-frame selected sample such as ours.

\section{\label{dearth} On the dearth of $z>15$ galaxies}

\begin{figure}[h!]
    \includegraphics[width=\columnwidth]{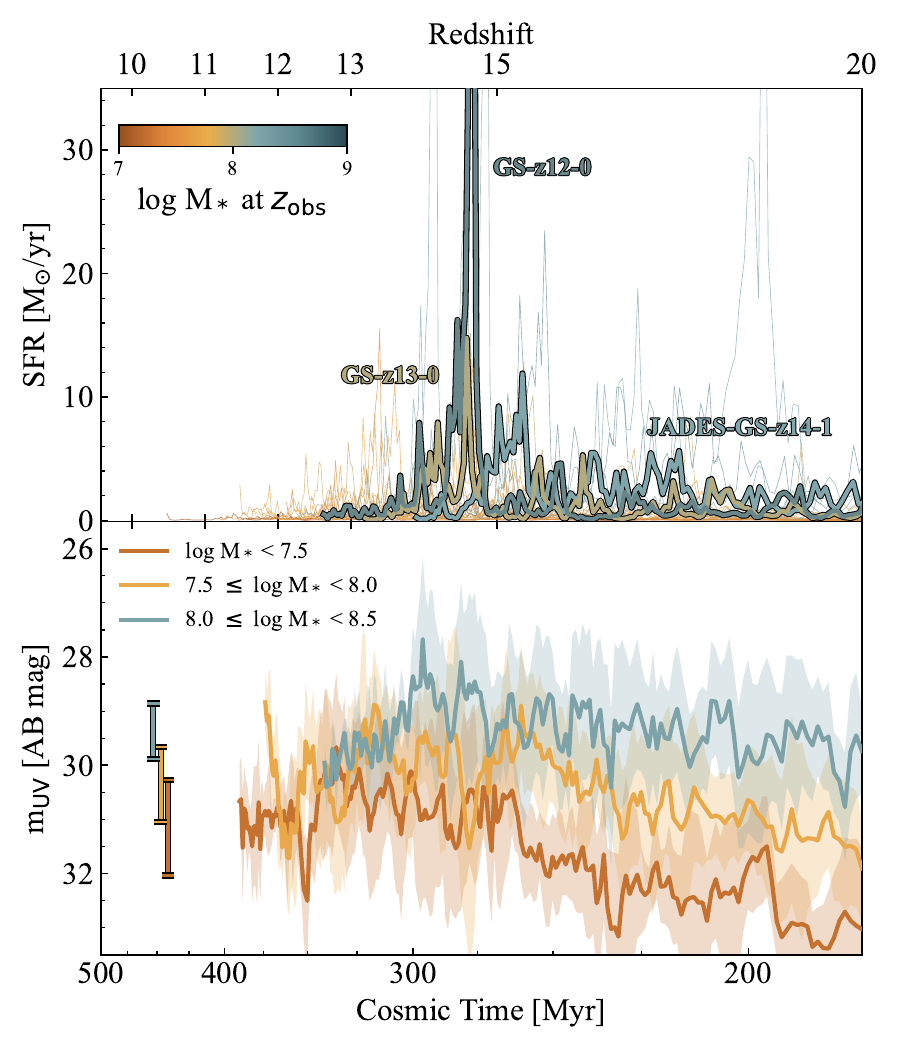}
 \caption{SFH of the galaxies that have reached 20\% of their stellar mass by $z=15$, that is 28 galaxies from our sample. Top panel: Lines are color-coded according to the stellar mass of the galaxies at $z_{\rm obs}$. Thick lines are the SFH of the three spectroscopically confirmed sources, well-known from the literature: GS-z12-0, GS-z13-0, and GS-z14-1 \citep{CurtisLake23,Carniani24}. Bottom panel: Median and standard deviation of apparent UV magnitude as a function of redshift in three different stellar mass bins. The vertical bars indicate the standard deviation of the median m$_{\rm UV}$ for each stellar mass bin.}
 \label{unicorn_sfh_z15+}
\end{figure}

In this section, we focus on the 28 galaxies of our sample that have formed at least 20\% of their stellar mass by $z=15$. We show their SFH in Fig.~\ref{unicorn_sfh_z15+}.
Among these 28 galaxies, three are known, spectroscopically confirmed sources in the GOODS-South field: GS-z12-0 and GS-z13-0 \citep{CurtisLake23}, and GS-z14-1 \citep{Carniani24}, we highlight their SFH on Fig.~\ref{unicorn_sfh_z15+}.
For these sources, we use the spectroscopic redshift for the SED fitting and obtain $\log(M_\star/M_\odot)=8.37\pm0.22$, $8.00\pm0.31$, and $8.22\pm0.26$, for GS-z12-0, GS-z13-0, and GS-z14-1, respectively.
\cite{CurtisLake23} obtained $\log(M_\star/M_\odot)=7.64^{+0.66}_{-0.33}$ for GS-z12-0, and $7.95^{+0.19}_{-0.29}$ for GS-z13-0, assuming a constant SFH.
For GS-z14-1, \cite{Carniani24} obtained $\log(M_\star/M_\odot)=8.0^{+0.4}_{-0.3}$.
Regarding the SFR, we find slightly lower values, with $\rm SFR=0.7\pm0.5$, $0.7\pm0.5$, and $0.4\pm0.3\,M_\odot\,{\rm yr}^{-1}$, for GS-z12-0, GS-z13-0, and GS-z14-1, respectively.
\cite{CurtisLake23} obtained higher values for GS-z12-0 and GS-z13-0, $1.8\pm0.5$ and $1.4^{+0.3}_{-0.2}\,M_\odot\,{\rm yr}^{-1}$, respectively. 
For GS-z14-1, \cite{Carniani24} estimated an $\rm SFR_{10}=2.0^{+0.7}_{-0.4}\,M_\odot\,{\rm yr}^{-1}$ while our estimate of the SFR averaged over the last 10\,Myr is $0.7\pm0.3\,M_\odot\,{\rm yr}^{-1}$.
We attribute these discrepancies to the different assumptions made in the SED modeling, particularly regarding the SFH, and to the fact that in this work we fit photometry only.

\begin{figure}[h!]
    \includegraphics[width=\columnwidth]{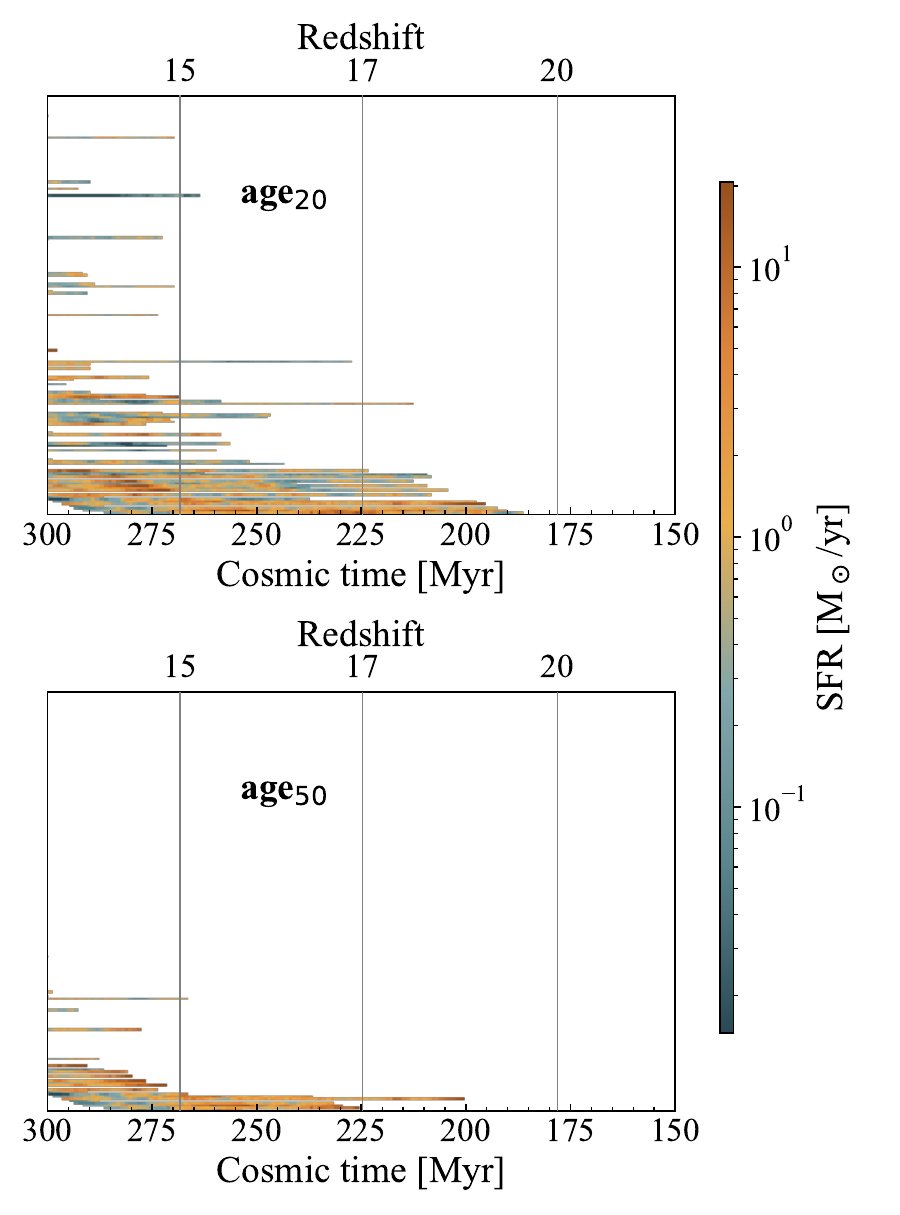}
 \caption{Zoom-in version of Fig.~\ref{unicorn_ridge} focusing on the galaxies that were forming stars at $z=15$. Top panel shows the zoom-in, starting the SFHs when the galaxies formed 20\% of their stellar mass while the bottom panel is the same but starting the SFH when the galaxies formed 50\% of the stellar mass. The y-axis scales are the same in both panels to help the comparison.}
 \label{unicorn_ridge_zoomin}
\end{figure}

The struggle to detect galaxies at $z\geq15$ with JWST has raised the question whether we are hitting a physical or observational barrier.
In Sect.~\ref{sec:sfh}, we showed that the vast majority of galaxies at $z\leq10$ did not form more than 20\% of their stellar mass by $z=12$ and do not seem to be the clear descendants from a population of galaxies that could have already formed by $z\sim15$.
On the other hand, some galaxies at $z\sim14$ have been spectroscopically confirmed \citep[e.g.,][]{Carniani24,RobertsBorsani24}.
If the $z=15$ barrier is physical, these galaxies should be no more than a few tens of Myr old. 
In Fig.~\ref{unicorn_ridge_zoomin}, we show a zoom-in of Fig.~\ref{unicorn_ridge} focusing on these galaxies that had already formed more than 20\% of their stars by $z=15$ (top panel).
For these 28 galaxies, the median redshift of formation, considering $age_{\rm 20}$, is $z_{\rm 20}=16.9^{+1.1}_{-1.1}$ for a median $age_{\rm 20}=118^{+14}_{-17}$\,Myr.
If we consider $age_{\rm 50}$, which is similar to the mass-weighted age (Fig.~\ref{agedef}), then the median formation redshift is $z_{\rm 50}=14.5^{+0.6}_{-0.7}$ for a median $age_{\rm 50}=64^{+12}_{-4}$\,Myr.
Even though some galaxies are old with one of them having even formed 20\% of its mass by $z=20$, statistically these numbers indicate that 75\% of the galaxies active at $z=15$ were active at $z<18.0$ and $z<15.1$, considering $age_{\rm 20}$ and $age_{\rm 50}$, respectively.
In Fig.~\ref{unicorn_sfh_z15+} we show the median SFH of these galaxies as well as the apparent UV magnitude derived from the conversion of the SFR for three stellar mass bins.
The expected m$_{\rm UV}$ are 31.2$\pm$0.9, 30.4$\pm$0.7, and 29.4$\pm$0.5, for $\log(M_\star/M_\odot)<7.5$,  $7.5<\log(M_\star/M_\odot)<8.0$, and $8.0<\log(M_\star/M_\odot)<8.5$, respectively.

\begin{figure}[h!]
    \includegraphics[width=\columnwidth]{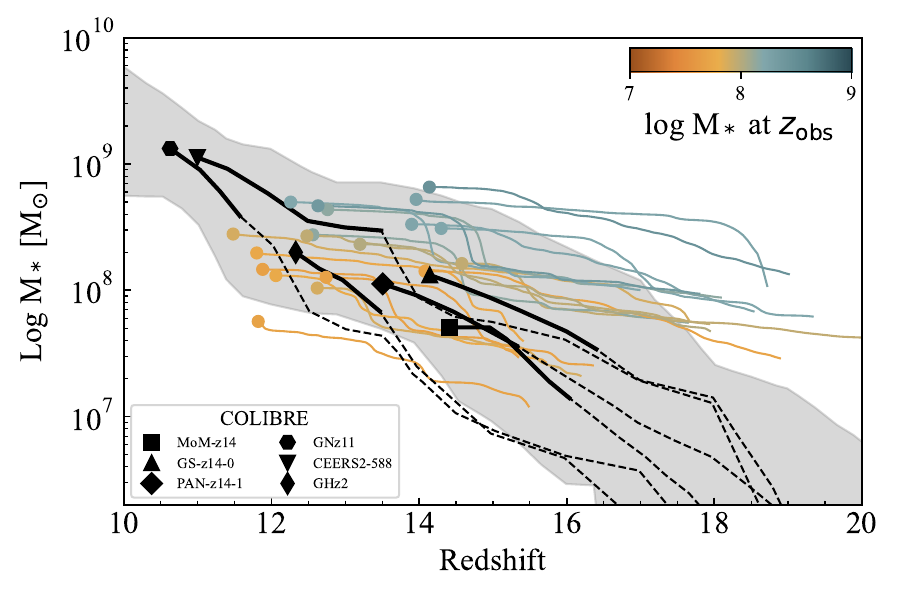}
 \caption{Stellar mass history of the progenitors of well-known $z>10$ galaxies from the \textsc{colibre} simulation (black symbols and lines) compared with the galaxies of our sample which have already formed 20\% of their stellar mass by $z=15$ and with similar stellar masses. The black solid lines are the stellar mass history of the progenitors in the simulation from the moment they reached 20\% of their stellar mass. The dashed lines show the mass history before this threshold. The boundaries of the grey region show the minimum and maximum values of the mass history of the progenitors of similar stellar mass in the \textsc{colibre} simulation. Our reconstructed mass histories show a slower growth compared to \textsc{colibre} but almost all of our SFH reconstructions are compatible with the range of mass histories predicted by \textsc{colibre}.}
 \label{unicorn_cum_z15}
\end{figure}

Recently, \cite{Chaikin26} studied the progenitors of well-known $z>10$ sources in the \textsc{colibre} simulations \citep{Schaye26}. 
They used the \textsc{colibre} L200m6 cosmological simulation, calibrated only on local-Universe galaxy properties, making its high-redshift predictions genuine tests of the model, to examine the most massive galaxies at $z>10$. 
They selected 515 galaxies with $>$10$^{8}$\,M$_{\odot}$ at $z=10$ and traced their main progenitors back to the onset of star formation, finding that these systems build up stellar mass extremely rapidly, with more massive $z=10$ galaxies tracking to higher masses and more massive host haloes at all earlier times. 
In the most extreme mass bin (10$^{9.5}$-10$^{10}$\,M$_{\odot}$), star formation begins as early as $z=20$, with stellar mass surging roughly three orders of magnitude, from $\sim$10$^{6.5}$ to $\sim$10$^{9.5}$\,M$_{\odot}$, in only about 300\,Myr, while their host haloes grow from $\sim$10$^{10.2}$-10$^{10.6}$\,M$_{\odot}$ at $z=14$ to 10$^{11.2}$\,M$_{\odot}$ by $z=10$.
In Fig.~\ref{unicorn_cum_z15}, we compare their stellar mass histories to the ones we reconstructed from our SED modeling. 
We include the \textsc{colibre} progenitors of specific well-known sources: MoM-z14 \citep{Naidu26}; GNz11 \citep{Oesch14}; GS-z14-0 \citep{Carniani24}; CEERS2-588 \citep{Finkelstein23}; PAN-z14-1 \citep{Williams25,Donnan26}; GHz2 \citep{Castellano22,Naidu22}, and limit our sample to similar stellar mass range.
Sixteen out of 19 of our reconstructed SFHs are within the global range of mass histories of the \textsc{colibre} simulation (grey region), showing a relatively good agreement at first order.
Our SFHs, however, tend to show a slower growth of the stellar mass and thus older $age_{\rm 20}$ while still within the global range of the \textsc{colibre} mass histories for the low-mass ones ($\log(M_\star/M_\odot)\lesssim8$).
The difference is more pronounced for our more massive galaxies but we note that at similar redshift ($z\sim14$), the galaxies tested by \cite{Chaikin26} are less massive, hence not exactly comparable.

\section{\label{backintime} Tracing back in time to $z\sim17$}

For the 28 sources that already have formed 20\% of their stellar mass by $z=15$, we use the reconstructed SFHs to recreate the SED and thus estimate the UV magnitude that they would have emitted at $z=17$ and $z=20$, assuming no dust attenuation, provided that they had already formed 20\% of their stellar mass at these redshifts.
Eleven galaxies satisfy this criterion at $z=17$ and only one at $z=20$.
For the eleven sources at $z=17$, we obtain a median M$_{\rm UV}$ of $-18.9^{+0.6}_{-0.5}$\,mag.
For an individual SFH, the median variation caused by the stochasticity of the SF activity over the last 10\,Myr is 1.5\,mag; we use this as an error on M$_{\rm UV}$ in our tests.
Varying the metallicity or the escape fraction marginally changed the distribution of UV magnitudes, hence we assume a metallicity of 0.004 and an escape fraction of 10\%.

\begin{figure}[h!]
    \includegraphics[width=\columnwidth]{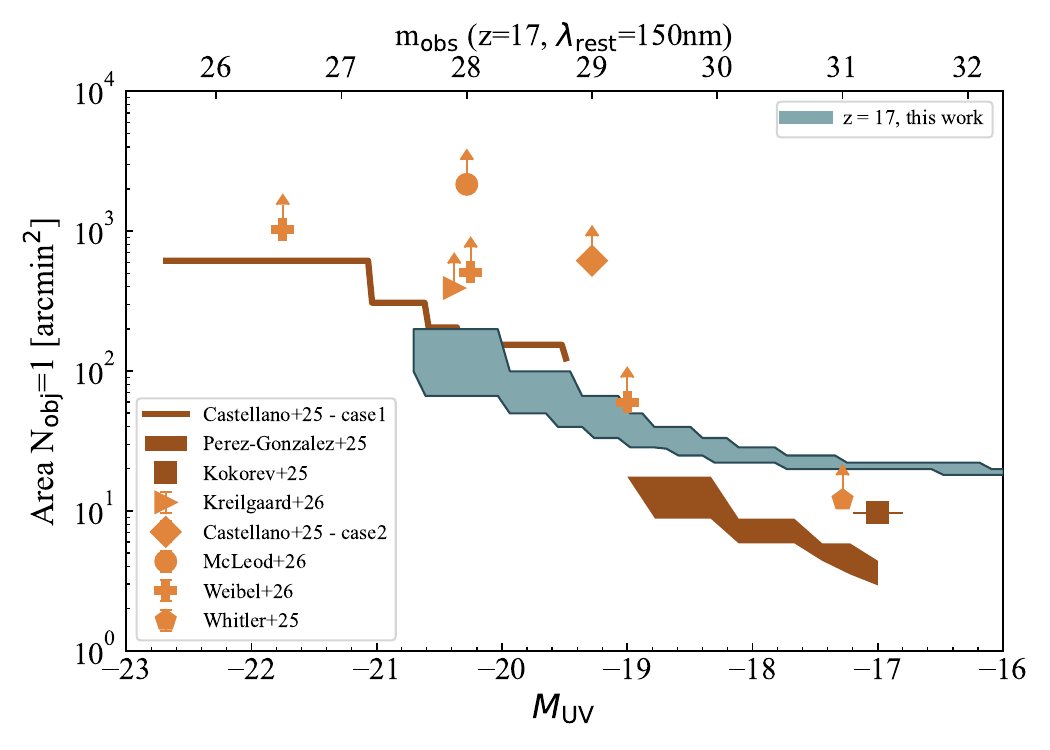}
 \caption{
Area at which at least one candidate brighter than $M_{\rm UV}$ at $z\sim17$ is detected. Figure adapted from \cite{Castellano25} and \cite{Fontana26}. The blue region shows the results from this work at different depth. Dark orange symbols and lines are $z\gtrsim15$ candidates while light orange ones are lower limits from the literature. 
 }
 \label{unicorn_castellano}
\end{figure}

Inspired by the figure proposed by \cite{Castellano25} and discussed in \cite{Fontana26}, we plot in Fig.~\ref{unicorn_castellano} the area to detect at least one object at $z\sim17$ as a function of UV magnitude.
We consider the 28 sources that already have formed 20\% of their stellar mass by $z=15$ and use their reconstructed UV magnitudes.
We take into account the impact of the burstiness of the SFH over the last 10\,Myr as an error, and show them in Fig.~\ref{unicorn_castellano}.
We compare our results with studies from the literature that searched for galaxies at $z>15$: \cite{Castellano25} using the ASTRODEEP-JWST covering several extragalactic fields \citep{Merlin24}; \cite{PerezGonzalez25} using MIRI Deep Imaging Survey and NGDEEP; \cite{Kokorev25} using GLIMPSE; \cite{Kreilgaard26} using BEACON;  \cite{McLeod26} from a compilation of extragalactic fields reaching 0.6\,sq.deg; \cite{Weibel26} with the PANORAMIC survey; and \cite{Whitler25} using JADES.
The results of these studies are divided into two groups: the first one did not find any candidate of galaxies at $z\sim17$ \citep{McLeod26,Weibel26,Whitler25,Kreilgaard26} or test the case where their candidates are interlopers \citep{Castellano25}.
The consequence of these results is a rapid decline of the UVLF at $z>15$ and a drastic drop in galaxy number counts happening between $z=14$ and $z=17$, thus in 60-65\,Myr.
The other analyses \citep{Castellano25,PerezGonzalez25,Kokorev25} have proposed $z\sim17$ candidates that, if confirmed, would point towards a slower evolution of the UVLF with redshift.
Our recovered $z\sim17$ UV magnitudes are compatible with, or slightly more optimistic than \cite{Castellano25} at the bright end (M$_{\rm UV}\lesssim-19.5$) but we predict more sources at fixed area than the other studies.
At fainter magnitude, we predict fewer sources than \cite{PerezGonzalez25} at fixed area by a factor $\sim3-8$ at M$_{\rm UV}\gtrsim-18$\,mag.
Still at the faint end, our predictions are closer to \cite{Kokorev25} (less than a factor of 2) and compatible with the upper limit of \cite{Whitler25} who used the JADES field to look for candidates, same field as we are using in this study.
Our reconstructed SFHs thus point toward a relatively slow decline of star formation between $z=14$ and $z=17$.
However, out of the eleven sources already forming stars at $z=17$, only one of them was active at $z=20$.
This is very low statistics and subject to the uncertainties of SED fitting and SFH reconstruction, but it would lead to a tentative estimate of Area$_{\rm Nobj=1}\geq$200\,arcmin$^2$ at M$_{\rm UV}=-18.1\pm0.6$.
This would be a factor 8 difference with our prediction at $z=17$ at the same UV magnitude, hence pointing toward a rapid evolution between $z=17$ and $z=20$.

\section{\label{conclusions}Conclusions}

In this study, we have proposed a new SFH model based on Ornstein-Uhlenbeck process to simulate bursty star formation histories of galaxies in the early Universe.
The burstiness level of the SFH can be modulated so that it varies with time.
We tested this model and its ability to reconstruct the SFH of $z>8$ galaxies on \textsc{sphinx$^{20}$} mock catalogs.
The goal is not to reproduce each star formation burst of the SFH, but rather to propose a model that provides good physical properties estimates along with an SFH close to what is expected in high-$z$ galaxies, that is with bursty star formation events and lull phases.
We fit the photometric mock catalogs released as part of the \textsc{sphinx$^{20}$} deliverables, we do not use spectroscopic information.
The results of this test show that:
\begin{itemize}
    \item The model provides good estimates of the stellar masses and SFR$_{10}$ of the \textsc{sphinx$^{20}$} mock galaxies.
    \item At population level, the model recovers well the SFHs of the \textsc{sphinx$^{20}$} galaxies, as probed by their cumulative star formation histories.
    \item The model struggles to recover the formation redshift, $z_{\rm form,20\%}$, or equivalently age$_{\rm 20}$ for galaxies at $z=8$. For the oldest galaxies at $z=9$ and the full mock sample at $z=10$, the formation redshift/age is well recovered.
\end{itemize}
This new SFH model is public and available through the \textsc{cigale} \textsc{sfhbursty} module\footnote{\url{https://gitlab.lam.fr/cigale/cigale/-/commit/421b9d47c7be97d5384350bcd0cf204b553b27c6}}.

Once the SFH model had been tested and calibrated, we use it on the UNICORN/GOODS-South sample v0.91 and fit 400 galaxies at $9\leq z \leq 15$. The redshift range is chosen to have a good reconstruction of the SFH and good constraints on the formation redshift.
From the SED modeling and SFH reconstruction, we find that:
\begin{itemize}
    \item Galaxies at $z\sim9$ reached 20\% of their stellar mass not earlier than z$\sim$12, suggesting that they may not be the descendant of galaxies observed and spectroscopically confirmed at $z\gtrsim13$.
    \item Out of the 218 galaxies of our sample that are at $10<z<15$, only 28 have formed 20\% of their stellar mass before $z=15$ and only one at $z=20$.
    \item Using as time of birth the age of the galaxies when they reached 20\% of their stellar mass, the median formation redshift of these 28 galaxies is $z_{\rm form,20\%}=16.9^{+1.1}_{-1.1}$. Changing the age definition using 50\% as the threshold, the median formation redshift is $z_{\rm form,50\%}=14.5^{+0.6}_{-0.7}$. In other words, 75\% of the galaxies actively forming stars at $z=15$ had formed 20\% of their stellar mass no earlier than $z=18$.
    \item For these galaxies, the expected median m$_{\rm UV}$, computed over the full SFH of the galaxies, are 31.2$\pm$0.9, 30.4$\pm$0.7, and 29.4$\pm$0.5, for $\log(M_\star/M_\odot)<7.5$,  $7.5<\log(M_\star/M_\odot)<8.0$, and $8.0<\log(M_\star/M_\odot)<8.5$, respectively.
    \item At $z=15$, 24\% of our sample has already formed 10$^7$\,M$_{\odot}$. Their median redshift is $z=10.9^{+0.9}_{-0.5}$, confirming though that the bulk of the galaxies at $z=9$ are unlikely the descendants of the $z>13-14$ galaxies.
    \item Our reconstructed SFHs are in agreement with the predictions of the mass growth of the \textsc{colibre} simulation in a similar stellar mass range, although the recovered SFHs from this work show a slightly slower growth compared to the simulation's predictions.
\end{itemize}

We then used the reconstructed SFHs to go back in time and estimate the UV emission of the galaxies at $z=17$.
Only 11 out of the 28 galaxies forming stars at $z=15$ were active, by our definition, at $z=17$.
We determine the area needed to have at least one galaxy at $z=17$ as a function of UV magnitude.
For bright sources (M$_{\rm UV}<-19.5$), our predictions are consistent with, or slightly more optimistic than, the results of \cite{Castellano25}.
In the faint part (M$_{\rm UV}>-18$), our results are consistent with the constraint from \cite{Whitler25} and indicate that a wider area is needed compared to the constraints from \cite{PerezGonzalez25} and \cite{Kokorev25}.
At $z=20$, only one source is found to be forming stars from its reconstructed SFH, putting a tentative constraint (or possibly lower limit) of $\sim$200\,arcmin$^2$ at M$_{\rm UV}=-18.1\pm0.6$.
These results point toward a relatively slow evolution of star formation activity between $z\sim15$ and $z\sim17$ with a more rapid decline between $z\sim17$ and $z\sim20$.
We note, however, that the analysis is subject to observational constraints that bias the sample towards the most luminous sources, and to the typical uncertainties linked to SED modeling.
Indeed, our sample is selected at $9<z<15$, and is biased towards galaxies that have already assembled sufficient stellar mass to be detectable at these redshifts. 
Reconstructing their progenitors at $z=17$ or $z=20$ does not provide a complete or unbiased census of the galaxy population at these epochs. 
For instance, galaxies that were already forming stars at $z=17$, but had not yet become sufficiently massive or luminous to enter our observed sample at later times, are not accounted for in our analysis.
Nevertheless, it proposes a different approach on the redshift frontier debate.
Further surveys with JWST/NIRCam and/or MIRI will add further constraints on the evolution of the UVLF at $z>15$ and the onset of star formation.

\begin{acknowledgements}
This work received support from the French government under the France 2030 investment plan, as part of the Initiative d’Excellence d’Aix-Marseille Université – A*MIDEX AMX-22-RE-AB-101.
LC and OI acknowledge the funding of the French Agence Nationale de la Recherche for the project iMAGE (grant ANR-22-CE31-0007).
MB acknowledges support by the ANID BASAL project FB210003. This work was supported by the French government through the France 2030 investment plan managed by the National Research Agency (ANR), as part of the Initiative of Excellence of Université Côte d’Azur under reference No. ANR-15-IDEX-01. This research was funded, in whole or in part, by the French National Research Agency (ANR), grant ANR-25-CE31-0192 (project CIGALE2D).
NIRCam was built by a team at the University of Arizona (UofA) and Lockheed Martin's Advanced Technology Center, led by Prof. Marcia Rieke at UoA.
This work is based on observations made with the NASA/ESA/CSA James Webb Space Telescope. 
The data were obtained from the Mikulski Archive for Space Telescopes at the Space Telescope Science Institute, which is operated by the Association of Universities for Research in Astronomy, Inc., under NASA contract NAS 5-03127 for JWST. 
JADES DR5 includes NIRCam data from JWST programs 1176, 1180, 1181, 1210, 1264, 1283, 1286, 1287, 1895, 1963, 2079, 2198, 2514, 2516, 2674, 3215, 3577, 3990, 4540, 4762, 5398, 5997, 6434, 6511, and 6541.
The authors acknowledge the teams of programs 1895, 1963, 2079, 2514, 3215, 3577, 3990, 6434, and 6541 for developing their observing program with a zero-exclusive-access period.
\end{acknowledgements}

\bibliographystyle{aa}
\bibliography{paper_SFH_transition.bib}

\end{document}